 \documentstyle[ emulateapj,flushrt,epsf,rotate]{article}

\def\bt{\begin{tabbing}}
\def\et{\end{tabbing}}

\def\beq#1{\begin{equation}\label{#1}}
\def\eeq{\end{equation}}
\def\beqa#1{\begin{eqnarray}\label{#1}}
\def\eeqa{\end{eqnarray}}
\def\eq#1{equation~(\ref{#1})}
\def\Eq#1{Equation~(\ref{#1})}

\def\xbf {{\bf x}}
\def\Xbf {{\bf X}}

\def\ybf {{\bf y}}
\def\ybft{{\bf y}^T}
\def\nbf {{\bf n}}

\def\Cbf {{\bf C}}
\def\Lbf {{\bf L}}
\def\Lbft{{\bf L}^T}
\def\zbf {{\bf z}}
\def\zbft{{\bf z}^T}
\def\Ibf {{\bf I}}
\def\rbf   {{\bf r}}
\def\rbfHat{\hat{\rbf}}

\def\dT  {\Delta T}

\def\albf    {{\bf a}}
\def\albfHat {\widehat{{\bf a}}} 
\def\da      {\Delta\albfHat}

\def\syn{synchrotron~}

\def\microk{$\mu$K~}
\def\mmicrok{$\mu$K}
\def\microm{$\mu$m~}
\def\um{\mu{\rm m}}

\def\mj{MJy/sr}

\def\lletter{$Letter$}

\def\etal{{\frenchspacing\it et al. \frenchspacing}}
\def\eg{{\frenchspacing\it e.g.}}
\def\ie{{\frenchspacing\it i.e.}}
\def\l{\ell}

\def\ith{i^{th}}
\def\jth{j^{th}}
\def\expec#1{\langle#1\rangle}
\def\SS{{\bf\Sigma}}

\def\spose#1{\hbox to 0pt{#1\hss}}
\def\simlt{\mathrel{\spose{\lower 3pt\hbox{$\mathchar"218$}}
     \raise 2.0pt\hbox{$\mathchar"13C$}}}
\def\simgt{\mathrel{\spose{\lower 3pt\hbox{$\mathchar"218$}}
     \raise 2.0pt\hbox{$\mathchar"13E$}}}
\def\simpropto{\mathrel{\spose{\lower 3pt\hbox{$\mathchar"218$}}
     \raise 2.0pt\hbox{$\propto$}}}

\def\pp{\noindent\parshape 2 0truecm 13.6truecm 1truecm 12.6truecm}
\def\rn{\pp}

\def\bfig{\begin{figure}[h] \centerline{\hbox{}}\vfill}
\def\efig{\end{figure}\vfill\newpage}

 \journalid{337}{15 January 1989}
 \articleid{11}{14}

\lefthead {DE OLIVEIRA-COSTA {\etal}}
\righthead{DIFFUSE GALACTIC EMISSION}

\begin{document}

\title{Cross-correlation of Tenerife data with Galactic 
       templates --- evidence for spinning dust?}

\author{
        Ang\'elica de Oliveira-Costa$^{1,2}$, 
                         Max Tegmark$^{1,3}$, 
	         Carlos M. Gutierrez$^{4}$,
	               Aled W. Jones$^{5}$, 
		        R. D. Davies$^{6}$, 
	               A. N. Lasenby$^{5}$, 
		           R. Rebolo$^{4}$ \&
	              R. A. Watson$^{6,4}$} 
  

\begin{abstract}
The recent discovery of dust-correlated diffuse microwave emission 
has prompted two rival explanations: free-free emission and 
spinning dust grains. We present new detections of this component 
at 10 and 15 GHz by the switched-beam Tenerife experiment. The data show
a turnover in the spectrum and thereby supports
the spinning dust hypothesis. We also present a significant detection
of synchrotron radiation at 10 GHz, useful for normalizing 
foreground contamination of CMB experiments at 
high-galactic latitudes.
\end{abstract}



\section{INTRODUCTION}

Understanding the diffuse microwave emission from the Galaxy 
is crucial for doing cosmology with Cosmic Microwave Background 
(CMB) anisotropies. 
Although three components of Galactic emission have 
been firmly identified (\syn and free-free radiation, and thermal 
emission from dust particles), it is important to better quantify
their frequency dependence and spatial distribution
(see, \eg, Kogut \etal 1996a; Tegmark \etal 1999 and references therein). 

Cross-correlations of CMB data with far-IR maps have shown the 
existence of a microwave emission component whose spatial 
distribution is traced by these maps (Kogut \etal 1996b; 
Lim \etal 1996; de Oliveira-Costa \etal 1997; Leitch \etal 1997;
de Oliveira-Costa \etal 1998, hereafter dOC98). Although this emission 
component has a spectral index suggestive of free-free 
emission (Kogut 1999),
the correlations between H$\alpha$ and dust maps have been found to be 
marginal (McCullough 1997; Kogut 1997). The source of this 
correlated emission is therefore an open question. Recent work suggests that 
it originates from spinning dust grain emission 
(Draine and Lazarian 1998), which should have a spectral index of 
$-3.3 < \beta_{spin} < -4$ between 19 and 53~GHz. Since spinning 
dust grains have a predicted turndown in their emission at frequencies 
below 20~GHz, a cross-correlation analysis with lower frequency measurements 
may help to discriminate between free-free and spinning dust emission models.

The purpose of this letter is to evaluate the Galactic contribution 
in the Tenerife 10 and 15~GHz data sets by cross-correlating them with 
the DIRBE dust maps and with the Haslam and Reich and Reich \syn maps.

\bigskip
{\small{
\noindent
$^{1}$Institute for Advanced Study, Olden Lane, Princeton, NJ 08540; angelica@ias.edu 

\noindent
$^{2}$Princeton University, Dept. of Physics, Princeton, NJ 08544

\noindent
$^{3}$Hubble Fellow 

\noindent
$^{4}$Instituto de Astrofisica de Canarias, 38200 La Laguna, Tenerife, Spain

\noindent
$^{5}$Mullard Radio Astronomy Observatory, Cavendish Laboratory, Madingley Road, Cambridge, CB3 0HE, UK

\noindent
$^{6}$University of Manchester, Nuffield Radio Astronomy Laboratories, Jodrell Bank, Macclesfield, Cheshire, SK11 9DL, UK
}}

\noindent
Our analysis is based on the Tenerife  
measurements made up until the end of 
1997 (Guti\'errez \etal 1999, hereafter G99).
The Tenerife switched-beam experiment is 
a sky survey between $0^{\circ} \le {\rm RA} \le 360^{\circ}$ 
and $30^{\circ} \le \delta \le 45^{\circ}$ 
(Figure~1 shows a sample strip at $\delta=32^\circ$) 
carried out at an angular 
resolution of $~ 5.1^{\circ}$ FWHM with an instrument that uses a 
double-differencing technique. Data was taken at frequencies 10 and 
15~GHz (we omit the 33~GHz data here due to lack of 
low latitude sky coverage), 
which are treated separately in order to gain additional frequency 
information on Galactic emission. 

\vspace{-2cm}
\centerline{{\vbox{\epsfxsize=9cm\epsfbox{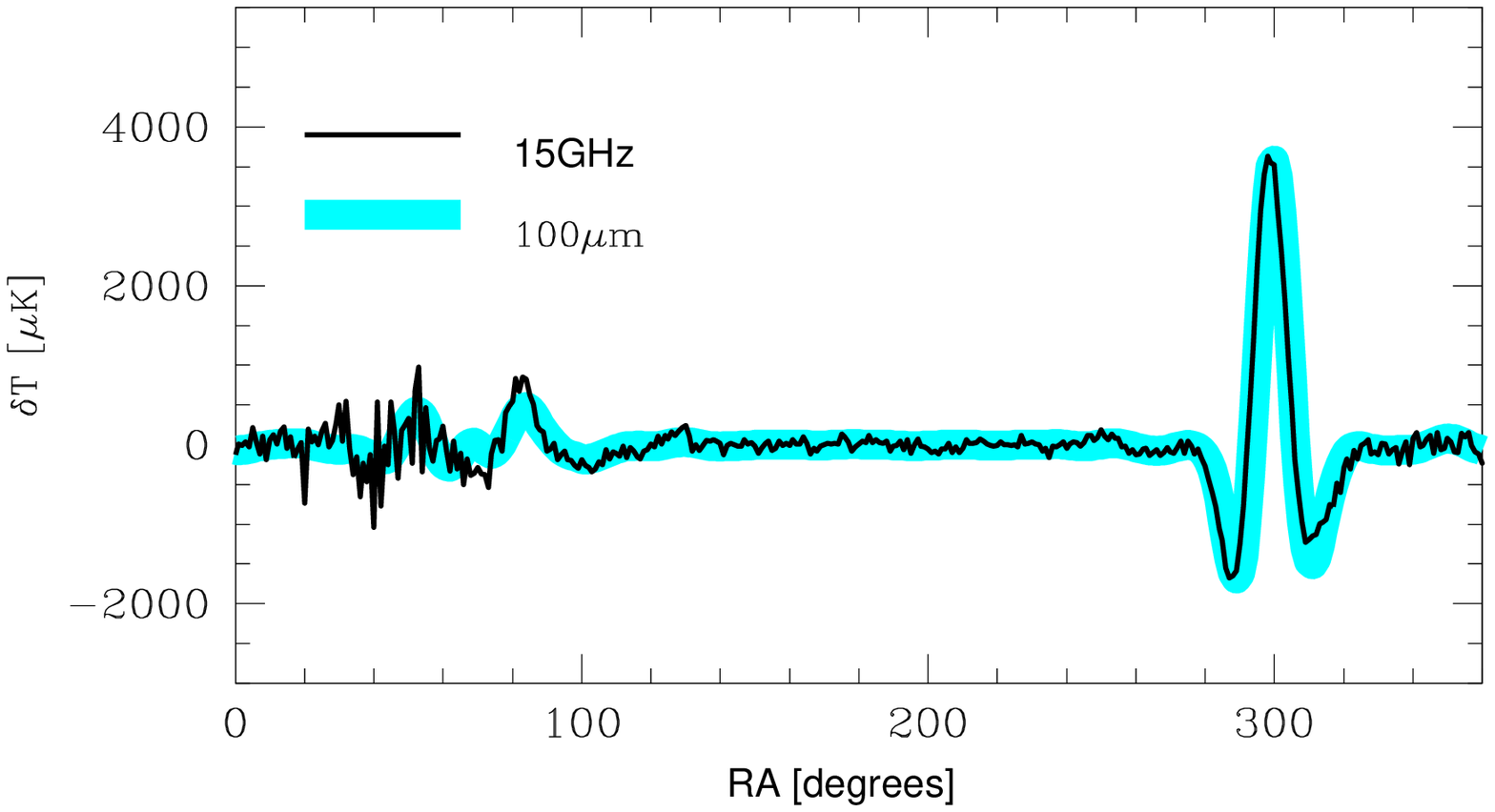}}}}
\vspace{-2cm}
\noindent{\small
 	Fig.~1 --- 
	Scan of the Tenerife region 
	$0^{\circ} < {\rm RA} < 360^{\circ}$ for $\delta = 32^{\circ}$.
	The thin curve shows Tenerife 15~GHz data, while
	the thicker curve represents DIRBE 100\microm convolved with
	the Tenerife triple-beam, rescaled for best fit. The spikes at 
        RA$\sim 80^{\circ}$ and $300^{\circ}$ correspond to the Galactic 
        plane crossings.
\label{figTen1}
}

\section{METHOD}

The Tenerife data consists of $N=2880$ pixels
at each frequency with 
double-differenced sky temperatures $y_i$ and noise $n_i$.  
We assume that this data is a linear superposition 
of CMB fluctuations $x_{CMB}^i$ and $M$ Galactic components 
whose angular distributions are traced in part by external 
foreground templates.
Writing these contributions as $N$-dimensional vectors, we obtain
\beq{signals}
	\ybf = \Xbf\albf + \xbf_{CMB} + \nbf,
\eeq
\goodbreak
\noindent
where $\Xbf$ is an $N\times M$ matrix whose rows
contain the various foreground templates convolved with the 
Tenerife triple-beam (\ie, $\Xbf_{ij}$ would be the $\ith$
observation if the sky had looked like the $\jth$ foreground 
template), and $\albf$ is a vector of size $M$ that gives the 
levels at which these foreground templates are present in the 
Tenerife data.

We treat $\nbf$ and $\xbf_{CMB}$ as uncorrelated random vectors 
with zero mean, and the matrix $\Xbf$ as constant; thus the data 
covariance matrix is given by
\beq{varCMB}
  \Cbf \equiv 
       \expec{\ybf \ybf^T} - \expec{\ybf} \expec{\ybf^T} =
       \expec{\xbf_{CMB} \xbf_{CMB}^T} + \expec{\nbf \nbf^T}.
\eeq
The Tenerife noise covariance matrix $\expec{\nbf \nbf^T}$
is, to good approximation, diagonal (see Guti\'errez de la Cruz \etal 1995),
while the CMB covariance matrix has the form 
\beq{CMBcovarEq}
  \expec{\xbf_{CMB} \xbf_{CMB}^T}_{ij} = 
  	\sum_{k=-1}^1 \sum_{l=-1}^1 w_k w_l ~ 
  	c(\rbfHat_{ik} \cdot \rbfHat_{jl}),
\eeq
where the vector ${\bf w}\equiv(-0.5,1,-0.5)$ gives the weights 
of the three lobes in the Tenerife triple beam, $\rbfHat_{i0}$ is 
the direction that the central lobe pointed to during the $\ith$ 
observation, and $\rbfHat_{i,\pm 1}$ is at the same declination but  
offset by $\pm 8.1^\circ$ on the sky. We take the CMB correlation function 
to be
\beq{corrFunc}
	c(\rbfHat_{ik} \cdot \rbfHat_{jl}) = 
	c(\cos\theta) 			   \equiv 
	\sum_{\l=2}^\infty B_\l^2 C_\l P_\l(\cos\theta)
\eeq
for a flat power spectrum $C_\l\propto 1/\l(\l+1)$ 
normalized to a Q$_{flat}=(5C_2/4\pi)^{1/2}=20${\mmicrok}
(G99). Finally, 
we model the Tenerife beam as a Fisher function 
(Fisher {\etal} 1987) 
\beq{rot2}
	B(\rbfHat\cdot\rbfHat') = 
	\frac{\exp(\rbfHat \cdot\rbfHat' / \sigma^2)}
	{4 \pi \sigma^2\sinh (\sigma^{-2})}
\eeq
with FWHM$=\sqrt{8\ln 2}\sigma=5.1^\circ$, which gives a window 
function $B_\l\approx e^{-\sigma^2\l(\l+1)/2}$.
The Fisher function is locally Gaussian and integrates to 
unity over the unit sphere.

Since our goal is to measure $\albf$, 
both $\nbf$ and $\xbf_{CMB}$ act as unwanted noise
in \eq{signals}.
Minimizing
$ \chi^2 \equiv 
          (\ybf - \Xbf\albf)^T 
          {\bf C}^{-1}
          (\ybf - \Xbf\albf) $
yields the minimum-variance estimate of $\albf$,
\beq{alpha}
   \albfHat  = 
   \left[\Xbf^T ~ {\bf C}^{-1} ~  \Xbf\right]^{-1}
   \Xbf^T ~ {\bf C}^{-1} ~  \ybf,
\eeq
with covariance matrix
\beq{varalpha}
   {\SS}\equiv\expec{\albfHat^2} - \expec{\albfHat}^2 =
   \left[\Xbf^T ~ {\bf C}^{-1} ~ \Xbf\right]^{-1}.
\eeq
The error bars on individual correlations are therefore 
$\Delta \widehat{a}_i=\SS_{ii}^{1/2}$. This includes the effect of chance 
alignments between the CMB and the various template maps, since 
the CMB anisotropy term is incorporated in 
$\expec{\xbf_{CMB} \xbf_{CMB}^T}$.

\section{DATA ANALYSIS \& RESULTS}

Due to the double-differencing technique, the Tenerife data are 
insensitive to the monopole ($\ell=0$) and dipole ($\ell=1$). 
Accordingly, when we convolve the template maps with the Tenerife 
triple beam function, we are removing the mean of the templates
as well as large angular scale structure. As a consequence, our 
results depend predominantly on the small scale intensity variations 
in the templates ($\ell\sim 10-30$) and are insensitive to the zero 
levels and gradients in the CMB and the template maps. 

\bigskip
{\footnotesize\center{Table~1. -- Correlations for 10 and 15~GHz data.}
 \vspace{-0.1cm}
\begin{center}
\begin{tabular}{llrrrr}	
\hline
\multicolumn{1}{c}{$b$ \& $\nu$}	  &	
\multicolumn{1}{c}{Template$^{(a)}$}      &	
\multicolumn{1}{r}{$\albfHat\pm\da^{(b)}$}& 
\multicolumn{1}{r}{${\albfHat\over\da}  $}&    
\multicolumn{1}{c}{$\sigma_{Gal}        $}&          
\multicolumn{1}{c}{$\dT$ [\mmicrok]$^{(c)}$}\\	
\hline
\hline
$|b| > 20^{\circ}$	&100\microm  &   49.8$\pm$11.2 &{\bf 4.5 }& 0.8   &   38.1$\pm$ 8.5 \\
10 GHz			&Has         &   29.5$\pm$ 5.9 &{\bf 5.0 }& 1.0   &   30.8$\pm$ 6.1 \\
 			&R\&R        &    0.8$\pm$ 0.3 &{\bf 2.8 }&23.2   &   17.8$\pm$ 6.4 \\
\hline
			&100\microm  &   71.8$\pm$ 4.5 &{\bf 5.9 }& 0.7   &   52.8$\pm$ 3.3 \\
15 GHz			&Has         &    0.8$\pm$ 3.4 &   0.2    & 1.1   &    0.8$\pm$ 3.7 \\
			&R\&R        &    0.1$\pm$ 0.2 &   0.3    &24.6   &    1.3$\pm$ 3.7 \\
\hline
\hline
$|b| > 30^{\circ}$	&100\microm  & $-$8.3$\pm$31.9 &$-$0.3    & 0.2   & $-$1.9$\pm$ 7.2 \\
10 GHz			&Has         &   35.8$\pm$ 8.8 &{\bf 4.1 }& 0.9   &   31.1$\pm$ 7.6 \\
			&R\&R        &    0.5$\pm$ 0.4 &   1.2    &19.4   &    9.3$\pm$ 7.8 \\
\hline
			&100\microm  &   94.9$\pm$15.3 &{\bf 6.2 }& 0.3   &   24.9$\pm$ 4.0 \\
15 GHz			&Has         & $-$8.3$\pm$ 5.0 &$-$1.7    & 0.9   & $-$7.3$\pm$ 4.4 \\
			&R\&R        & $-$0.3$\pm$ 0.2 &$-$1.4    &19.8   & $-$6.0$\pm$ 4.4 \\
\hline
\hline
$|b| > 40^{\circ}$	&100\microm  &   84.6$\pm$54.8 &   1.5    & 0.2   &   14.6$\pm$ 9.6 \\
10 GHz			&Has         &   24.1$\pm$11.4 &{\bf 2.1 }& 0.8   &   19.4$\pm$ 9.1 \\
			&R\&R        &    0.4$\pm$ 0.5 &   0.6    &19.1   &    6.7$\pm$10.5 \\
\hline
			&100\microm  &   72.4$\pm$33.3 &{\bf 2.2 }& 0.2   &   12.3$\pm$ 5.7 \\
15 GHz			&Has         &$-$10.8$\pm$ 6.3 &$-$1.7    & 0.8   & $-$8.7$\pm$ 5.1 \\
			&R\&R        & $-$0.4$\pm$ 0.3 &$-$1.1    &19.5   & $-$7.0$\pm$ 6.2 \\
\hline
\end{tabular}
\end{center}
}
\vspace{-0.1cm}
\noindent{\small  
$^{(a)}$ The DIRBE and Haslam correlations listed in this table correspond to 
         joint 100$\um-$Has fits, whereas the R\&R numbers correspond to a joint 
         100$\um-$R\&R fit. \\
$^{(b)}$$\albfHat$ has units \microk (\mj)$^{-1}$ for the 100\microm template, 
         \mmicrok/K for the Has template and \mmicrok/mK for the R\&R template.
	 We use antenna temperature throughout. \\ 
$^{(c)}$ $\dT \equiv (\albfHat\pm\da) \sigma_{Gal}$.}
\bigskip

\subsection{Correlations \& Variances}\label{corrANDvar}

We cross-correlate the Tenerife data with two different \syn 
templates: the 408~MHz survey (Haslam \etal 1981) 
and the 1420~MHz 
survey (Reich and Reich 1988), hereafter Has and R\&R,
respectively. To study dust and/or free-free emission, we 
cross-correlate the Tenerife data with three Diffuse 
Infrared Background Experiment (DIRBE) sky maps at 
wavelengths 100, 140 and 
240~$\um$ (Boggess \etal 1992). The extent of point source 
contamination in the Tenerife data is discussed and estimated 
in G99, and therefore will not be addressed in 
this \lletter. In practice, we just remove the estimated point sources
contribution before calculating the correlations. 
For definiteness, we use the DIRBE 100$\um$ channel when placing all 
limits below since it is the least noisy of the three DIRBE channels, 
and the Haslam map since it is the \syn template at lowest frequency.

Table~1 shows the coefficients $\albfHat$ and the corresponding 
fluctuations in antenna temperature in the Tenerife data 
($\Delta T = \albfHat \sigma_{Gal}$, where $\sigma_{Gal}$ is 
the standard deviation of the template map).
The analysis is done for three different cuts: $20^{\circ},
30^{\circ}$ and the {\it Tenerife cut}
	(which consists of data with 
	$160^{\circ} < {\rm RA} < 250^{\circ}$, corresponding
	to Galactic latitudes $|b| \simgt 40^{\circ}$).
Note that the fits are done jointly for $M=2$ templates.
The DIRBE and Haslam correlations listed in Table~1 correspond to 
joint 100$\um-$Has fits, whereas the
R\&R numbers correspond to a joint 
100$\um-$R\&R fit.

Statistically significant ($>2\sigma$) correlations are listed in
boldface. 
The two \syn templates are found to be correlated with the 10~GHz 
data at low Galactic latitudes and this correlation persists with
the Has template even at higher latitudes, while the 15~GHz 
data is not correlated with these templates even for a $20^{\circ}$ 
cut. The 100$\um$ and 10~GHz data are correlated only at 
lower Galactic latitudes ($20^{\circ}$ cut), while the correlation 
with the 15~GHz data persists to higher latitudes. The same analysis 
was carried out for the DIRBE 140$\um$ and 240$\um$ maps, giving very 
similar results. 

\subsection{Latitude Dependence}

To investigate the dependence of the correlation on Galactic latitude,
we sliced the maps into six regions of equal area, each corresponding
to a range of latitude $|b|$. Figure~2 shows the results for the
100$\um$ map.
Note that $\albfHat$ from the 100$\um$$-$15~GHz correlation is
similar in all latitude bands, suggesting that the particular
ISM properties that are responsible for the correlation
do not vary strongly with latitude.

\vspace{-2.0cm}
\centerline{{\vbox{\epsfxsize=9cm\epsfbox{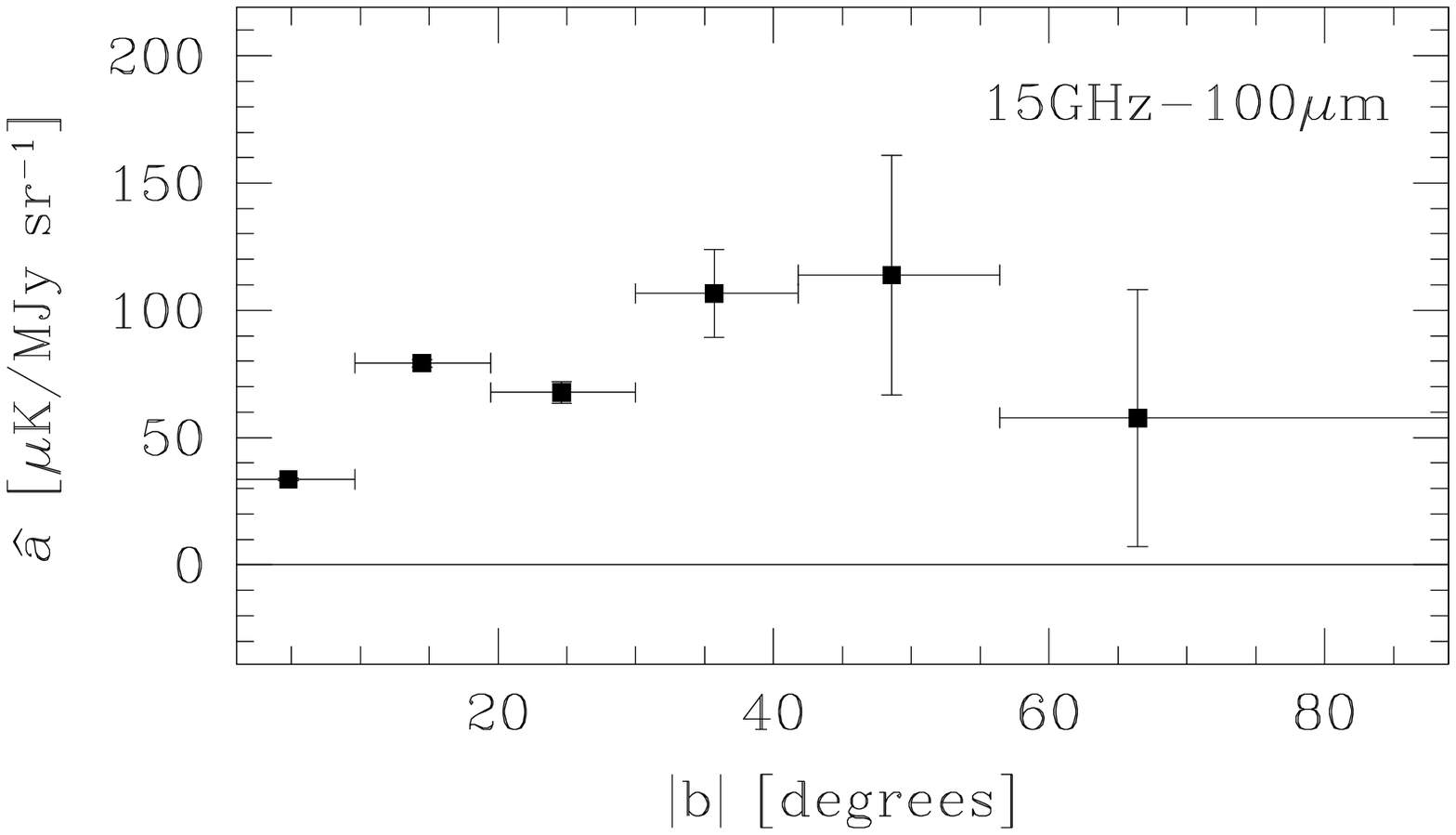}}}}
\vspace{-1.5cm}
\noindent{\small
 	Fig.~2 --- 
	Dependence of $\albfHat$ on Galactic latitude
	for the 100$\um$$-$correlated emission at 15 GHz.
\label{figTen2}
}
\bigskip

\subsection{Spectral Index}

Writing the frequency dependence as $\delta T\propto\nu^{\beta}$,
we obtain interesting limits on the spectral index $\beta$. 
For synchrotron radiation, both the Has$-$10~GHz
and the R\&R$-$10~GHz correlations are consistent with
$\beta\sim -3$ in all latitude slices.
The most favored values between 408~MHz and 10~GHz values 
($-3.2 \simlt \beta \simlt -3.4$) 
are slightly steeper than 
the canonical sub-GHz slope of 
$-2.7\simlt \beta \simlt -2.9$
(Davies {\etal} 1998; Platania {\etal} 1998). Such a steepening
of the spectrum at higher frequencies is
consistent with a steepening of the spectrum of 
cosmic ray electrons at higher energies
(Rybicki and Lightman 1979).
For the emission component correlated with
the 100$\mu$m map, the 
spectral index between 10 and 15~GHz is
$0.80^{+.77}_{-.66}$ 
for $|b| > 20^{\circ}$, and the spectrum
is also clearly rising at
$|b| \simgt 30^{\circ}$ (see Table~1). 
Calibration uncertainties are below 5\% in the Tenerife data
at both 10 and 15 GHz, and therefore do not affect the conclusions 
that we will draw from this spectral behavior.

\subsection{Effect of Template Correlations}\label{multifrequencyfit}

\Eq{varalpha} shows that we can interpret $\SS^{-1}$ as 
measuring correlations between the various templates.
Table~2 shows the dimensionless correlation coefficients
  $r \equiv \SS^{-1}_{ij} (\SS^{-1}_{ii}\SS^{-1}_{jj})^{-0.5}$,
using the 10 GHz pixel weighting, and 
enables a number of conclusions to be drawn:
\begin{enumerate}
\item 
The three DIRBE maps are, as expected, highly correlated,
\ie, they trace the same warm dust emission;
 
\item 
The two \syn maps are less well correlated, 
with only $r^2\sim 50\%$ of the variance in 
the Has map being traced by the R\&R map at 
$|b| > 20^{\circ}$ cut, and even less at 
higher latitudes\footnote{
       Systematic effects (such as 
       striping) are present in the \syn radio maps and may be 
       partially responsible for the uncorrelated 
       part of their signal (Davies {\etal} 1998). In addition, 
       detector noise in the template maps will 
       of course also reduce their correlation with each other and
       with the Tenerife data, but is in our case 
       negligible compared to the Tenerife noise.}; and
 
\item 
All three DIRBE maps are seen to be almost 
uncorrelated with both radio maps, with 
the common variance $r^2$ being only a few percent.
This means that $\SS$ is almost diagonal, and that
simply performing a separate correlation analysis
for each component (with $N=1$, as was done in 
\eg, de Oliveira-Costa {\etal} 1997; dOC98) will give 
almost identical results to a joint ($N>1$) analysis.
We tested this, and indeed obtained results in good agreement
with those presented in subsection \ref{corrANDvar}).
\end{enumerate}

\vspace{-0.1cm}
{\footnotesize\center{Table~2. -- Correlation between foreground 
                     templates$^{(a)}$.}
\vspace{-0.1cm}
\begin{center}
\begin{tabular}{lcccccc}	
\hline
	   & &100\microm &140\microm &240\microm &Has     &R\&R \\
&&&&&&\\	
100\microm & &1          &0.97       &0.96       &0.12    &0.14 \\      
140\microm & &0.91       &1          &0.98       &0.08    &0.11 \\ 
240\microm & &0.91       &0.88       &1          &0.07    &0.11 \\
Has 	   & &0.11       &0.12       &0.11       &1       &0.71 \\
R\&R 	   & &0.15       &0.11       &0.17       &0.53    &1    \\ 
\hline
\end{tabular}
\end{center}
}
\vspace{-0.1cm}
\noindent{\small $^{(a)}$ The upper right triangle of the table uses 
                 a $|b| > 20^{\circ}$ cut and the lower left triangle 
                 a $|b| > 30^{\circ}$ cut.} 
\medskip

\subsection{Fake Skies}\label{fakeskies}

We tested our software by analyzing 
constrained realizations of the CMB and the Tenerife instrument 
noise. Cholesky decomposing the covariance matrix as $\Cbf = \Lbf \Lbft$, 
we generated fake skies using
	$\ybf = \Lbf \zbf + \Xbf\albf$,
where $\zbf$ is a vector of independent Gaussian random variables
with $\expec{\zbf}={\bf 0}$ and $\expec{\zbf\zbf^t}=\Ibf$ (the identity 
matrix), which gives
$\expec{\ybf \ybft}-\expec{\ybf}\expec{\ybf}^t =
	 \Lbf \langle \zbf \zbft \rangle \Lbft =
	 \Lbf \Ibf    \Lbft   = \Cbf$.
Analyzing 1000 such realizations, we recovered unbiased 
estimates $\albfHat$ with a variance in agreement 
with \eq{varalpha}. 

\subsection{Sky Rotations}

Systematic errors or unmodeled foregrounds may add non-Gaussian
fluctuations to the the data set $\ybf$, making the true error bars
larger than suggested by \eq{varalpha}.
To address this issue, we repeated the fit done in \eq{alpha} after 
replacing the template map with a set of ``control'' patches selected 
from different portions in the sky. We repeated the analysis with 
24 $\times$ 2 $\times$ 2 = 96 transformed maps, rotated around the 
Galactic axis by multiples of 15$^{\circ}$ and/or flipped vertically 
and/or horizontally. 
The correct Haslam template has the highest of all 96 correlations with 
the 10~GHz data. 
Even for a $|b| \simgt 40^{\circ}$ cut, 95 of the 96 patches (or 99\% 
of them) are less correlated with the the 10~GHz data than the correct 
100$\um$ patch. 
The same test was carried out for other Galactic cuts 
and the 15~GHz data, giving similar results.  
These results show that our correlations are not due to systematic errors 
or chance alignments between the CMB and the various template maps. 
We also applied the rotation test to the foreground template correlations
in Table~2. Although as mentioned above, the 
dust-synchrotron correlations are tiny ($r\ll 1$) and negligible
for our purposes, they are still statistically significant;
for a 40$^{\circ}$ Galactic cut, 
the correct synchrotron template (Has or R\&R) 
has the highest of all 96 correlations with the DIRBE map.

\section{CONCLUSIONS}

We have detected high-latitude galactic foreground signals in the
Tenerife data at high ($\simgt 5\sigma$) statistical significance,
summarized in Figure~3.
The synchrotron signal is consistent with previous upper limits
(Kogut {\etal} 1996b) and detections (dOC98),
and provides a useful normalization for modeling microwave foregrounds.
We also verify that the 15 GHz {\it Tenerife-cut} data
is useful for CMB cosmology; although the DIRBE-correlated component 
is statistically significant, it
is much smaller in amplitude than the primary cosmological signal.

Our results indicate that the DIRBE-correlated signal 
may turn out to be the dominant foreground seen by the 
the MAP satellite.
What is its physical origin?
A ``smoking gun'' indication of spinning dust grains would be
a turnover in the spectrum, since all Draine and Lazarian models
show such a feature whereas the antenna temperature
of free-free emission continues to 
rise with $\beta\sim -2.1$ towards lower frequencies.
Figure~3 combines our present results with previous measurements.
The lower panel shows that the synchrotron spectrum is
well fit by a power law $\nu^\beta$ with $\beta\sim -3$ (cyan line).
In the upper panel, 
the dashed curve shows the spectrum for a best fit 
linear combination of vibrational (green) and rotational (yellow)
dust emission. The vibrational spectrum 
assumes a 20K dust temperature and an emissivity of 2.
The rotational spectrum is the warm interstellar medium model of
Draine and Lazarian (1998) from 
{\it www.astro.princeton.edu/$\sim$draine/}. 
A joint three-parameter fit, including free-free
emission with spectrum $\nu^{-2.15}$ as a third component (magenta), 
prefers negligible amounts of free-free above 10 GHz.
On purely physical grounds, free-free emission must of course 
be present at some level.
However, if the 10 GHz Tenerife correlation were entirely due to free-free emission,
this component would explain only about 10\%
of the variance observed at 15 GHz and 
a percent
of the variance in the Saskatoon data.
The interpretation of such fits that directly compare data 
points from different experiments is complicated by conversions 
between different angular scales (dOC98).
However, a direct comparison between the 10 and 15 GHz Tenerife points
from this work is straightforward, since the angular resolution and 
window functions are the same, and gives a spectral index of
$\beta=0.80^{+.77}_{-.66}$.
This is clearly inconsistent with free-free emission,
and provides evidence that the main culprit is indeed spinning dust.

\bigskip
The authors wish to thank 
Bruce Draine, Lyman Page, David Spergel and David Wilkinson 
for helpful comments.
Support for this work was provided by
NASA though grants NAG5-6034 and 
Hubble Fellowship HF-01084.01-96A (from STScI, operated by AURA, Inc. 
under NASA contract NAS5-26555), 
and NSF grant PHY-9600015.
AWJ acknowledges King's College, Cambridge, for
support in the form of a Research Fellowship. 

\medskip
\centerline{{\vbox{\epsfxsize=9cm\epsfbox{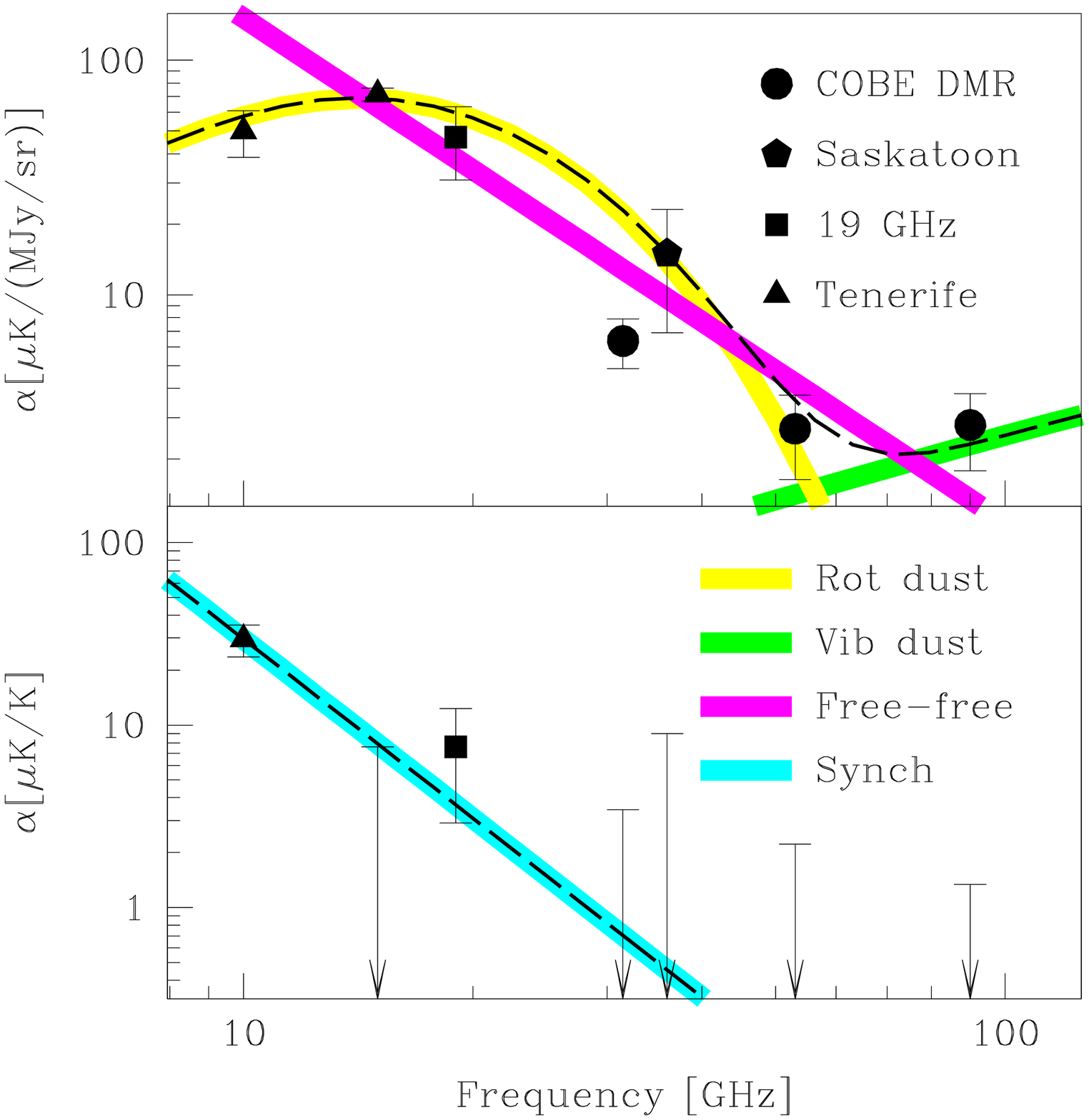}}}}
\noindent{\small
	 Fig.~3 --- 
	 Frequency dependence of 
	 DIRBE-correlated emission (top) and 
	 Haslam-correlated emission (bottom).
	 The DIRBE-correlated emission is seen to be
	 approximately fit by a combination of spinning and vibrating 
         dust (dashed curve), whereas free-free emission alone 
	 (falling straight line) 
	 cannot explain the drop from 15 to 10 GHz. 
	 The slope of the Haslam-correlated emission
	 is seen to be weekly constrained, fitting {e.g.} a single
	 $\beta=-3.26$ power law down to 408 MHz 
	 (where $\alpha=10^6\mu$K$/$K$=1$ by definition). 
	 Tenerife data is from Table 1
	 for a $20^\circ$ cut.
	 Upper limits are 2-$\sigma$. Note that the error bars on the 
         15 GHz point (top) are too small to be visible.
\label{figTen3}
}
\bigskip


\goodbreak

\end{document}